\documentclass[%
 aps,
superscriptaddress
 amsmath,amssymb,
 reprint,%
]{revtex4-2}

\usepackage{graphicx}
\usepackage{dcolumn}
\usepackage{bm}

\usepackage[utf8]{inputenc}
\usepackage[T1]{fontenc}
\usepackage{mathptmx}
\usepackage{etoolbox}
\usepackage{xcolor}
\usepackage{amsmath}
\usepackage{makecell}

\usepackage{listings}
\usepackage{xcolor}

\usepackage{float}

\makeatletter
\def\@email#1#2{%
 \endgroup
 \patchcmd{\titleblock@produce}
  {\frontmatter@RRAPformat}
  {\frontmatter@RRAPformat{\produce@RRAP{*#1\href{mailto:#2}{#2}}}\frontmatter@RRAPformat}
  {}{}
}%
\makeatother
\begin{document}

\preprint{AIP/123-QED}

\title{Machine learning analysis of structural data to predict electronic properties in near-surface InAs quantum wells}
\author{Patrick J. Strohbeen}
    \thanks{These authors contributed equally to this work}
\author{Abtin Abbaspour}
    \thanks{These authors contributed equally to this work}
\author{Amara Keita}
    \thanks{These authors contributed equally to this work}
\author{Tarek Nabih}
    \thanks{These authors contributed equally to this work}
\author{Aliona Lejuste}
\author{Andrea Maiani}
    \thanks{Now at NORDITA, Stockholm University}
\author{\\ Alisa Danilenko}
\author{Ido Levy}
\author{Jacob Issokson}
\author{Tyler Cowan}
\author{William M. Strickland}
\author{Mehdi Hatefipour}
\author{Ashley Argueta}
\author{Lukas Baker}
\author{Melissa Mikalsen}
\author{Javad Shabani}
\affiliation{ 
Center for Quantum Information Physics, Department of Physics, New York University, New York, NY 10003 USA
}%

\date{\today}

\begin{abstract}
Semiconductor crosshatch patterns in thin film heterostructures form as a result of strain relaxation processes and dislocation pile-ups during growth of lattice mismatched materials. Due to their connection with the internal misfit dislocation network, these crosshatch patterns are a complex fingerprint of internal strain relaxation and growth anisotropy. Therefore, this mesoscopic fingerprint not only describes the residual strain state of a near-surface quantum well, but also could provide an indicator of the quality of electron transport through the material. Here, we present a method utilizing computer vision and machine learning to analyze AFM crosshatch patterns that exhibits this correlation. Our analysis reveals optimized electron transport for moderate values of $\lambda$ (crosshatch wavelength) and $\epsilon$ (crosshatch height), roughly 1 $\mu$m and 4 nm, respectively, that define the average waveform of the pattern. Simulated 2D AFM crosshatch patterns are used to train a machine learning model to correlate the crosshatch patterns to dislocation density. Furthermore, this model is used to evaluate the experimental AFM images and predict a dislocation density based on the crosshatch waveform. Predicted dislocation density, experimental AFM crosshatch data, and experimental transport characterization are used to train a final model to predict 2D electron gas mean free path. This model shows electron scattering is strongly correlated with elastic effects (e.g. dislocation scattering) below 200 nm $\lambda_{MFP}$.
\end{abstract}

\maketitle

\section{Introduction}

One of the most common features of strained-layer epitaxy is the surface crosshatch pattern that forms as the crystal undergoes misfit dislocation formation, glide, and pile-up \cite{burmeister1969crosshatch, chang199035crosshatch}. However, as ubiquitous as these crosshatch patterns are, there are few studies attempting to bridge the gap between the observed surface morphologies and material parameters of interest, such as strain state or carrier scattering lifetimes \cite{andrews2002surfcrossstrain, fitzgerald1997epitaxystrain}. In principle, internal elastic strain and the surface crosshatch pattern that forms as a result are intimately related via the elasticity tensor of the material system. Proper definition of this interdependence will enable researchers to more rapidly characterize and optimize complex heterostructure growth. For example, an atomic force microscopy (AFM) image of a surface takes over an order of magnitude less time to complete when compared to cryogenic 4-point transport measurements. Previous models from the late 1990s and early 2000s attempted to solve this problem considering either a single uniform strain value \cite{fitzgerald1997epitaxystrain} or a single line of dislocations \cite{andrews2002surfcrossstrain}. However, expansion to two-dimensional surfaces and the use of these models to correlate surface corrugation to the quality of electronic transport has yet to be unveiled.

\begin{figure*}
    \centering
    \includegraphics[width=\linewidth]{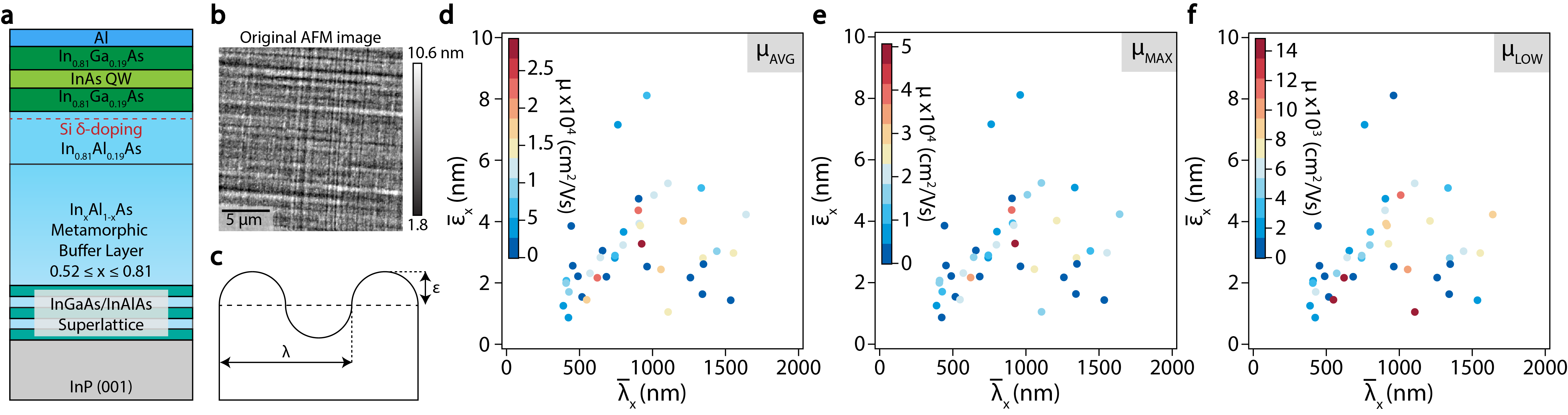}
    \caption{\textbf{a} Schematic of near-surface InAs quantum well heterostructure. \textbf{b} Exemplary 20x20 $\mu$m AFM image for a near-surface InAs quantum well sample JS629. \textbf{c} Schematic of crosshatch waveform model, adapted from \cite{fitzgerald1997epitaxystrain}. \textbf{d-f} Plots of quantum well mobility as a function of the average crosshatch amplitude ($\overline{\epsilon}$) and wavelength ($\overline{\lambda}$). }
    \label{afm}
\end{figure*}

Specifically, near-surface quantum well heterostructures within the last decade have been extensively studied for topological superconductivity \cite{shabani2016inas, frolov2020toposuper} and gate-tunable superconducting circuits for quantum information\cite{casparis2018gatemon, scappucci2021geinfo, hinderling2024geabsspec, tosato2023gehardgap, strickland2024gatemonloss}. These studies utilize material stacks in which the quantum well region is placed close to the surface of the wafer. In this case, the active region of the wafer is well-separated from the initial mismatched interface by the growth of substantial metamorphic buffer layers \cite{strickland2022inasqw, stehouwer2023gewafer}. Compared to buried quantum well structures which are well known to have carrier scattering dominated by background and ionized impurities \cite{chung2022gaaslimits}, modulation in the surface structure and surface dangling bonds have a much stronger impact on near-surface quantum well systems. As a result, understanding the \textit{surface} strain state and strain fluctuations becomes of greater importance. Therefore, in these heterostructures it is of great significance to develop models with which we can quickly understand and predict how these materials will perform in cryogenic applications without the need for long fabrication, cooldown, and measurement cycles.

Here, we present a method of AFM image processing for MBE-grown near-surface InAs quantum well structures in which we relate the surface crosshatch pattern to quantum well transport quality. The surface crosshatch pattern is a direct result of misfit dislocation formation during the growth process and is a sign of strain relaxation \cite{burmeister1969crosshatch, chang199035crosshatch}. Anisotropy in this pattern indicates anisotropic adatom diffusion along <110>-type directions in the (001) plane, which corresponds to anisotropic transport as well. We analyze the room temperature AFM images and extract a parametrization of the physical waveforms on the surface that define the crosshatch and corroborate this with the low temperature electron mean free path. By using the measured electron mean free path of the quantum well, we examine the relationship between the surface crosshatch and resulting transport properties in order to draw conclusions regarding quantum well transport behavior from surface morphology. Simulated 2D crosshatch patterns are used to correlate surface crosshatch patterns to an internal dislocation density for the experimental AFM images. The predicted dislocation densities are then used to model electron scattering mean free path values that are compared against the experimentally measured values. This study serves as a proof of concept for the applicability of computer analysis in the context of AFM image processing in which expanding to the utilization of machine learning is highly desirable.

\begin{figure}[h!]
    \centering
    \includegraphics[width=\linewidth]{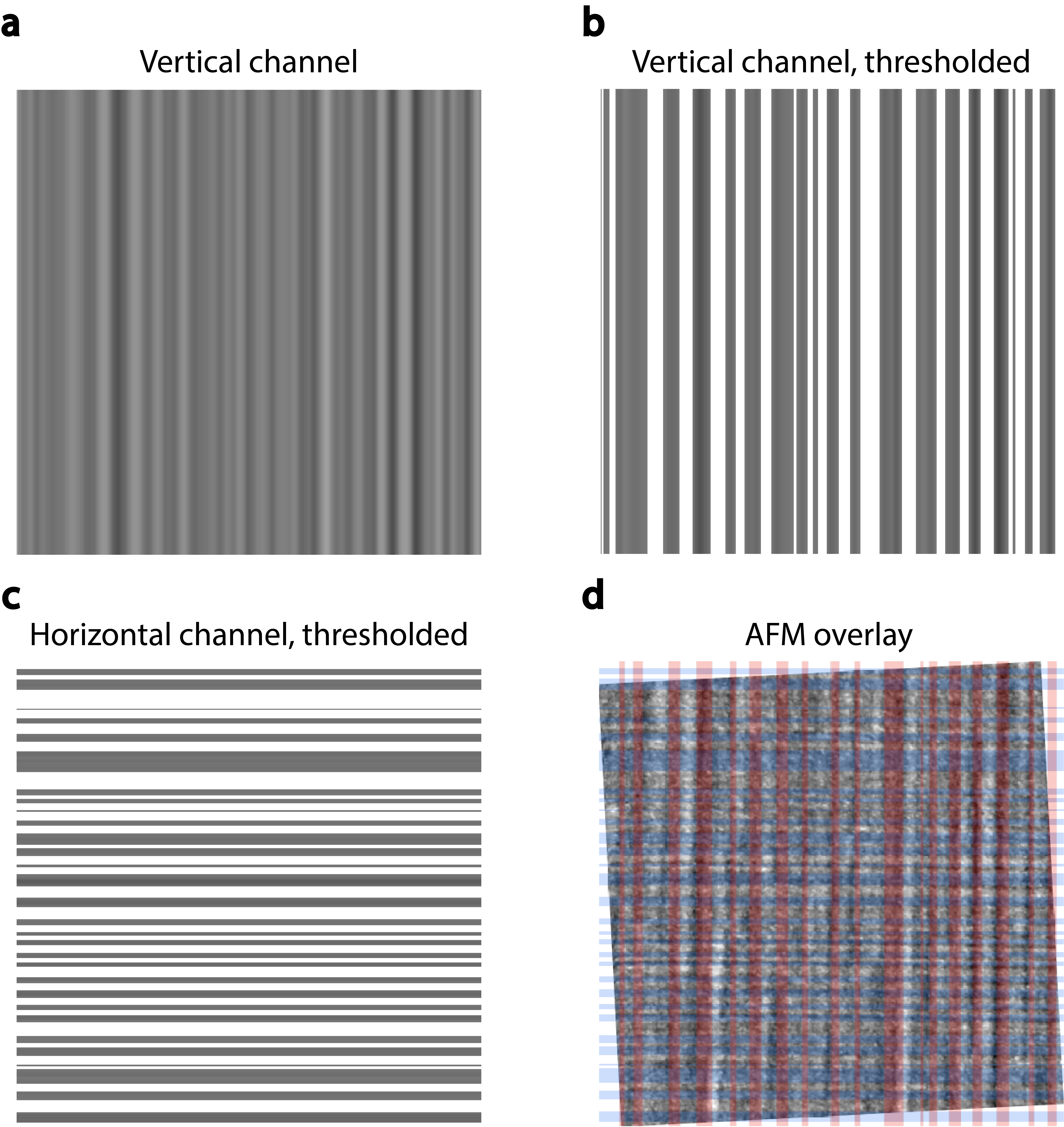}
    \caption{\textbf{Deconvolution of AFM image.} \textbf{a} Averaged color map for vertical crosshatch channel of sample JS629. \textbf{b} and \textbf{c} are the vertical and horizontal channels of sample JS629 after threshold processing to define the peaks and valleys of the waveform. \textbf{d} Overlaid vertical and horizontal channels from our AFM image processes algorithm on top of the original rotated AFM image from Fig. \ref{afm}b.}
    \label{afm_analysis}
\end{figure}

\begin{figure*}[ht!]
    \centering
    \includegraphics[width=\linewidth]{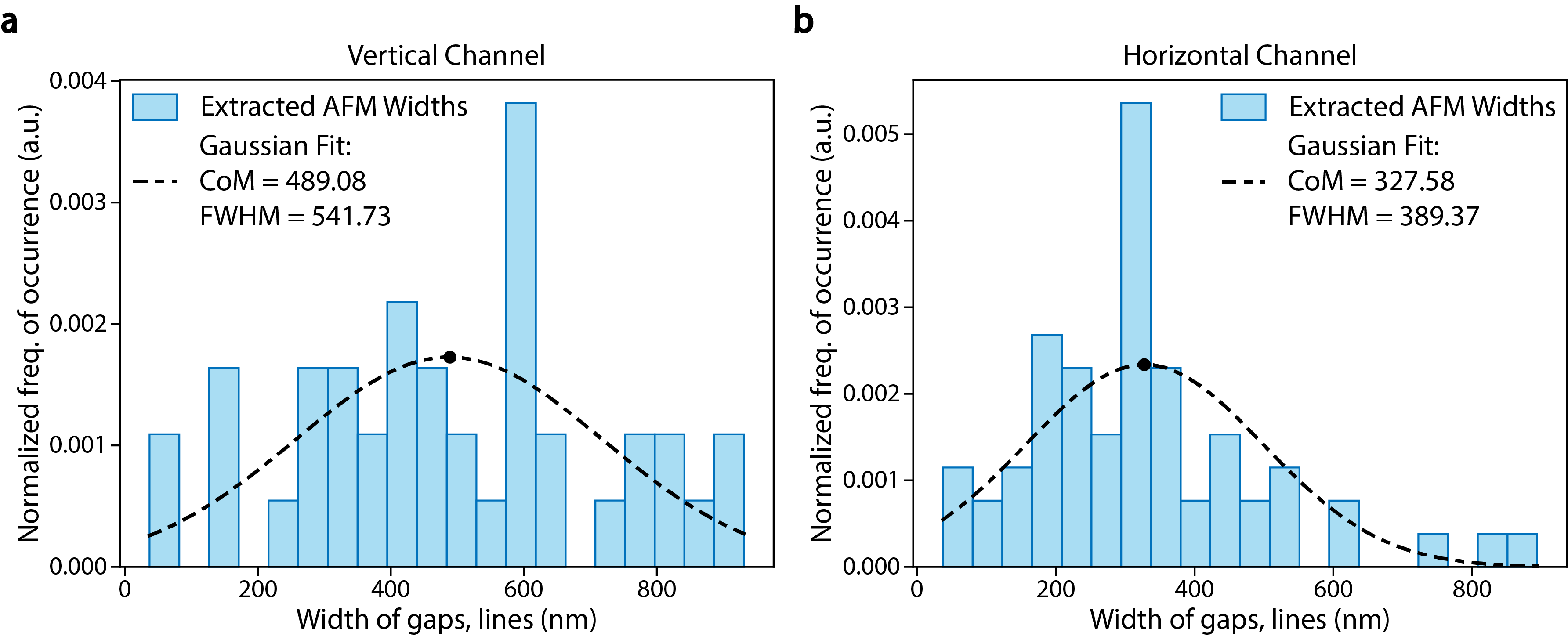}
    \caption{ \textbf{Extraction of $\lambda$.} Histogram plots in \textbf{a} and \textbf{b} for the vertical and horizontal channels, respectively, for the exemplary AFM image of sample JS629, presented in Fig. \ref{afm}a. The overlaid line in both plots is a Gaussian fit to the histogram from which we extract a center of mass value and a line width. These values are listed in the legend of this plot.}
    \label{afm_extract}
\end{figure*}

\section{AFM Crosshatch Extraction and Evaluation}

We first present our one-dimensional analysis of the crosshatch pattern. Assuming cubic symmetry of the surface, we would not expect drastically different behavior between the [110] and [1$\overline{1}$0] directions (parallel to the crosshatch peaks/valleys) within the (001) plane. Fig. \ref{afm} presents the InAs quantum well heterostructure studied here, an exemplary AFM input image, a schematic of our model definitions, and an example of a 1D analysis relating the electron mobilities as a function of the crosshatch wavelength and amplitude. The structure of the near-surface InAs quantum well is presented in Fig. \ref{afm}a, with an examplary input AFM image seen in Fig. \ref{afm}b. For our AFM crosshatch analysis we define the ``waveform'' of the crosshatch pattern in 1D. We define the wavelength, $\lambda$, as the distance between peaks in the crosshatch and the amplitude, $\epsilon$, is the height above the midpoint of the crosshatch. A schematic of this definition is presented in Fig. \ref{afm}c, which is an adaptation from the definition presented in Fitzgerald \textit{et al.} in 1997 \cite{fitzgerald1997epitaxystrain}. From the analysis we present in Figs. \ref{afm_analysis} and \ref{afm_extract}, we plot the 1D crosshatch definition presented here against quantum well 2D electron gas (2DEG) mobility for a series of 36 samples, seen in Fig. \ref{afm}d-f.  The $\mu_{MAX}$ and $\mu_{LOW}$ plots presented in Figs. \ref{afm}e and f, respectively, are the mobilities calculated for the $R_{xx}$ and $R_{yy}$ magnetoresistance traces. The average mobility in plot Fig. \ref{afm}d is the average for each sample as calculated from the high and low values seen in Fig. \ref{afm}e,f.

Simple intuition would lead one to believe that the optimal regime in terms of physical structure would be in the limit of $\overline{\epsilon} \rightarrow 0$ and $\overline{\lambda} \rightarrow \infty$. However, we note a general trend in our analysis suggesting an optimal point in the AFM crosshatch pattern in the region of moderate values of $\overline{\lambda}$ and $\overline{\epsilon}$. Thus, to more completely evaluate the AFM image in terms of electronic transport behavior, we extend our analysis software to incorporate the second dimension of the crosshatch pattern and fit our parametrization of the crosshatch to a linear regression model.

The AFM analysis is initialized by inputting the raw {20x20~$\mu$m} images and removing any residual borders from the normalization routines. The initial pixel dimension is then converted into physical length units (nm) based on the AFM scan size. After this step, AFM images are trimmed to standardized pixel densities. The AFM images are then rotated to align the surface crosshatch pattern such that the peaks and valleys are vertical (horizontal). To preserve as much as possible of the original data-set upon rotation, image resolution is increased, however, some data must be cut-off. For example, with a 512x512 pixel image we calculate a 4\% maximum possible loss depending on the degree of rotation.

After rotation, our AFM analysis software produces two images corresponding to the vertical (Y) and horizontal (X) channels and then applies a de-noising routine to enhance the signal-to-noise ratio (SNR). The de-noising procedure averages the pixel values in each column across the entire image, preserving the vertical crosshatch and condensing the original 2D image into an effectively 1D line. This line is then re-expanded to the original pixel size for the purpose of comparison to the original AFM image. This produces a smoothed version of the vertical crosshatch ``lines'', as seen in Fig. \ref{afm_analysis}a. This procedure is also done for the horizontal channel. We then apply a threshold such that height values are cut-off and replaced with empty pixels, examples of this for both the vertical and horizontal channels are shown in Fig. \ref{afm_analysis}b and c, respectively. The thresholding algorithm takes as an input the de-noised waveform that was created in the previous step and calculates the absolute midpoint between the maximum and minimum heights. Then, the local maxima (minima) are calculated for each peak (valley) above (below) the global midpoint. The threshold is applied to the image such that only information below the threshold value is kept. To check this result, we also present an overlay of the recreated crosshatch pattern on the original AFM image it was generated from in Fig. \ref{afm_analysis}d. In this image, we see good agreement between the ``valleys'' present in the crosshatch pattern and the red (vertical) and blue (horizontal) channels that our algorithm picks out.

Once the AFM images are processed and we obtain the thresholded channel plots as shown in Figs. \ref{afm_analysis}b and c, we then extract the widths of each line and gap between lines. These values are plotted in separate histogram plots for the vertical and horizontal channels, examples of which are shown in Figs. \ref{afm_extract}a and b. The histogram shows extracted values for the AFM image presented in Fig. \ref{afm}b and corresponds to the de-convoluted crosshatch graphs presented in Fig. \ref{afm_analysis}b and c. We then fit this histogram to a Gaussian distribution to extract a center of mass (CoM) value and the full-width at half max (FWHM). In the context of a real material evaluation, the CoM is related to the crosshatch density (residual strain state), whereas the FWHM value is then related to the fluctuations in this strain state.

To extract the height of the crosshatch, $\epsilon$, we use the height information from the AFM images to map the color values of each pixel back to the physical value. For each line, the peak-to-valley heights for the individual X- and Y-crosshatch patterns together to get $\epsilon_{x}$ and $\epsilon_{y}$, respectively. We define a mean value $\overline{\epsilon}_{x,y}$ across all 512 lines in the AFM image, from which we calculate the standard deviation, $\sigma^{\epsilon}_{x,y}$ for each AFM image. Thus, the resultant dataset for each AFM image contains $\lambda$, $\epsilon$, $\sigma^{\lambda}$, and $\sigma^{\epsilon}$ for X- and Y-channels for each sample analyzed in this method. 

\section{Linear Regression Model}

\begin{figure*}[htbp]
    \centering
    \includegraphics[width=0.9\linewidth]{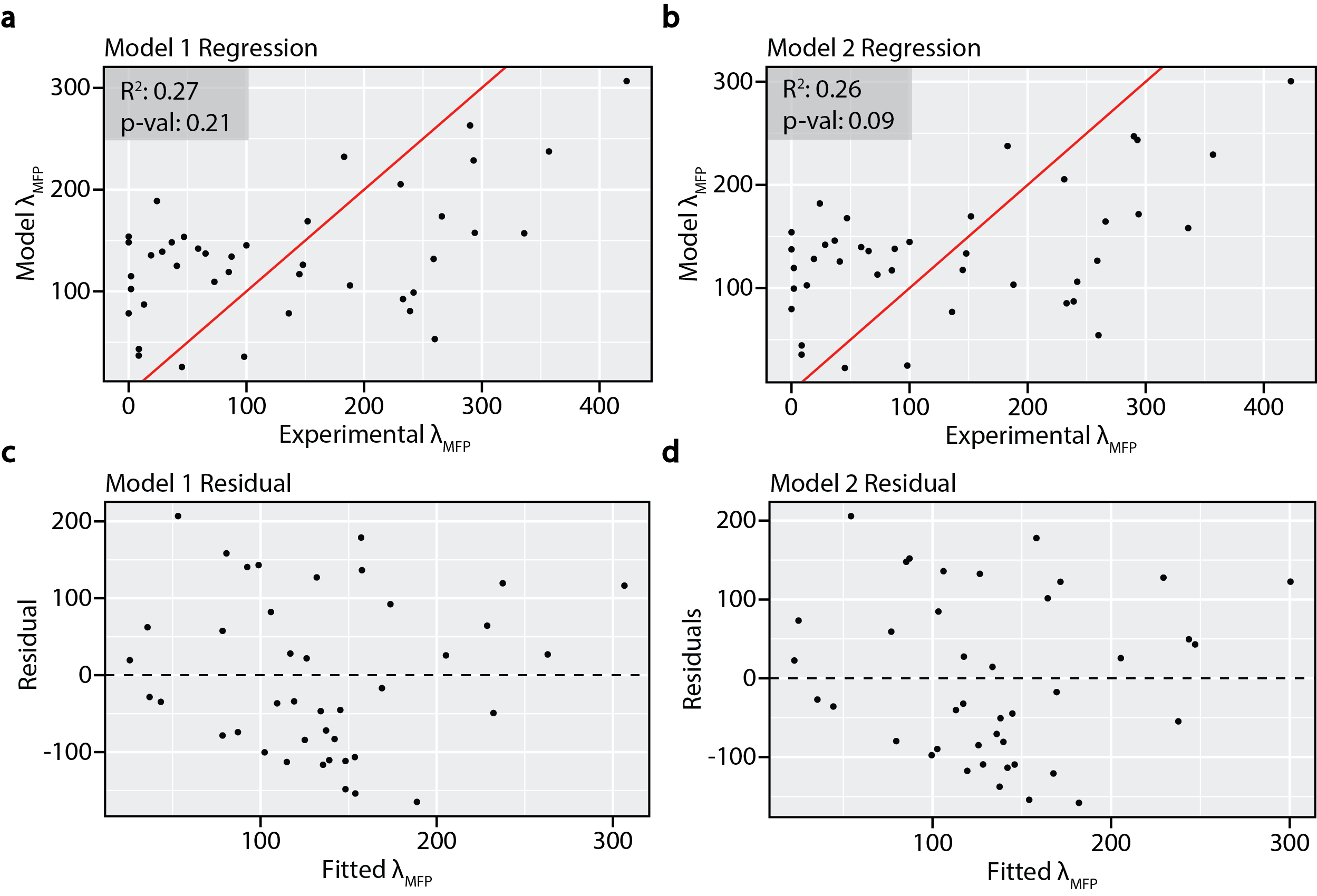}
    \caption{\textbf{Linear regression model fitting.} \textbf{a} Linear regression model fitting to the full dataset. The red line is the model fit. Calculated R$^{2}$ and p-values are stated in the inset. \textbf{b} Linear regression model fitting with $\lambda_{x}$ and $\lambda_{y}$ removed from dataset. The red line is the model fit. \textbf{c} Calculated residuals for model 1. \textbf{d} Calculated residuals for model 2.}
    \label{linear_regression}
\end{figure*}

We then fit the variables from the compiled dataset across all 36 samples to a multiple linear regression model using the \texttt{stats} package within base R statistical software (ver 4.3.3) \cite{rsoftware}. The results of two of these model fits are presented in Fig. \ref{linear_regression}. Figs. \ref{linear_regression}a,c are the regression model and calculated residuals for Model 1 and Figs. \ref{linear_regression}b,d are the same for Model 2. The difference between models is that for Model 2, the $\overline{\lambda}_{x}$ and $\overline{\lambda}_{y}$ variables are removed from the fit. The resulting linear regression equations for Model 1 and 2 are presented in the supplemental material. To generate the plots in Figs. \ref{linear_regression}a, b, the model $\overline{\lambda}_{MFP}$ value is calculated using the linear regression equations found in the supplmental material using the experimental dataset that is obtained through our computer vision algorithm. The $R^{2}$ and p-values for each model fit are presented in the insets in Figs. \ref{linear_regression}a and b. 

When considering all variables in our dataset (Model 1), we calculate the coefficient of determination, $R^{2}$, to be 0.27, with a statistical significance (p-value) of 0.21. This is suggestive of some relationship in the dataset to our desired transport parameter, though due to the complexity of the dataset the direct relationship is unclear. However, when removing the $\lambda$ values for both X- and Y-channels (Model 2), we calculate the $R^{2}$ value to be 0.26, i.e. roughly the same variance as in Model 1, except in the new model we calculate a p-value of 0.09. Thus, we conclude that there is a reasonably significant correlation between $\overline{\lambda}_{MFP}$ measured in transport and surface corrugation as defined by $\overline{\epsilon}_{x,y}$, $\sigma^{\lambda}_{x,y}$, and $\sigma^{\epsilon}_{x,y}$ (Model 2). The low $R^{2}$ value is likely due to our small sample size of only 36 samples. Furthermore, the improvement from the removal of the $\lambda_{x,y}$ variables from the linear model suggests that the dependance of transport quality on crosshatch wavelength is weak at best, but rather how non-uniform this crosshatch is. However, it is clear a more in-depth model is required to better capture the correlation between the corrugated crosshatch surface and the resulting electronic transport quality.

\section{2D Crosshatch Simulation}

As seen in Figure \ref{linear_regression}, a linear regression model fails to accurately capture the complexity of AFM crosshatch patterns. This suggests that higher dimensionality solutions may be necessary to realize the relationship between the elastic behavior of the semiconductors and their electronic transport behavior. To realize such an analysis, we develop simulation tools to simulate two-dimensional surface crosshatch patterns. This allows for the creation of a large machine learning model with which to better understand the physical system.

\begin{figure*}[htbp]
    \centering
    \includegraphics[width=\linewidth]{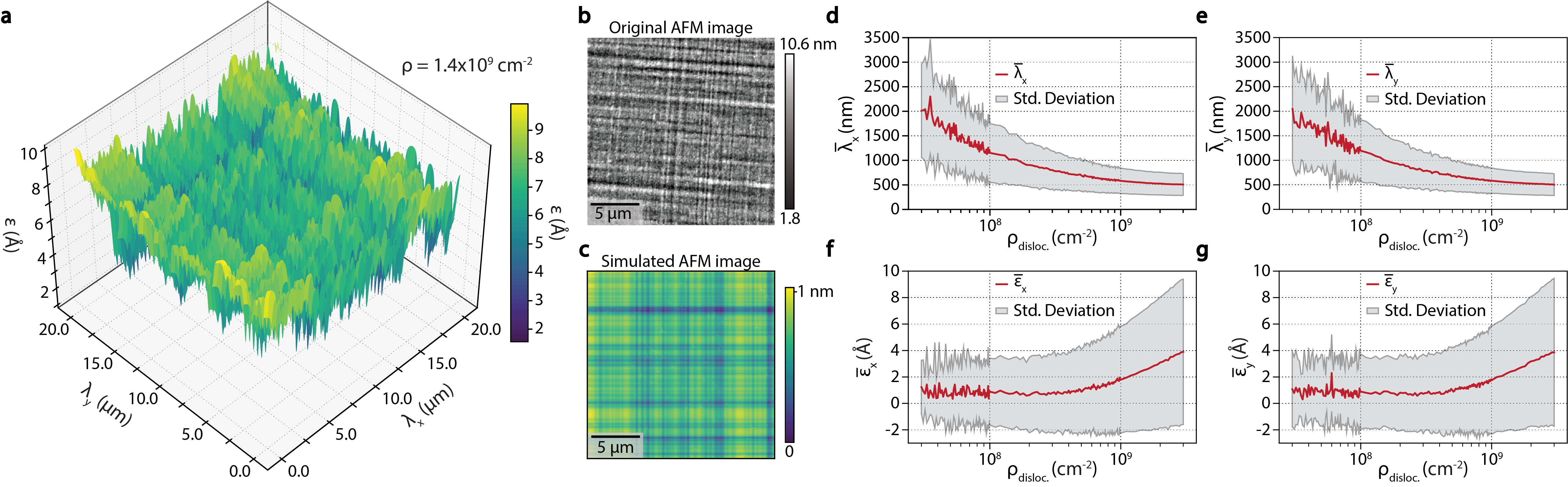}
    \caption{ \textbf{Crosshatch surface modeling.} \textbf{a} Simulated crosshatch surface for a starting density of $1.4 \times 10^{9}/cm^{2}$. \textbf{b} and \textbf{c} present exemplary experimental (``original'') and the 2D simulated AFM images, respectively, for comparison. The simulated image presented here is the 2D representation of the 3D crosshatch surface presented in \textbf{a}. \textbf{d} and \textbf{e}: $\overline{\lambda}_{x}$ and $\overline{\lambda}_{y}$, respectively, as a function of dislocation density from our crosshatch simulations. \textbf{f} and \textbf{g}: $\overline{\epsilon}_{x}$ and $\overline{\epsilon}_{y}$ as a function of dislocation density from our crosshatch simulations. In \textbf{d-g}, the red lines represent the average value across the repeat simulations for each dislocation density while the grey shaded area represents the first standard deviation.}
    \label{ml_model}
\end{figure*}

Our AFM simulation utilizes the model for surface crosshatch proposed by Andrews \textit{et al.} in 2002 \cite{andrews2002surfcrossstrain}. The original model solves Hooke's law in one dimension for an arbitrary elastic medium. The model is initialized by defining some starting density of dislocations along the a 1D line, and then integrating the strain field caused by the dislocations to a specified distance away from this starting point. This results in us obtaining an expression for the displacement field of the elastic medium at an arbitrary distance away. We first expanded the model to include two dimensions to generate a surface crosshatch in InAs from a given dislocation density. This model randomly places dislocations within a 20x20 $\mu$m region based on an initial user-defined dislocation density. The model equations are presented below in Equations \ref{x_displ} and \ref{y_displ} for calculating the displacement field in x and y, respectively.

\begin{equation}
    u_x = \frac{b}{\pi} \left[ \frac{h^2 - (x - x_0) \cdot h}{(x - x_0)^2 + (y - y_0)^2 + h^2} - \arctan\left(\frac{x - x_0}{h}\right) \right]
    \label{x_displ}
\end{equation}
\begin{equation}
    u_y = \frac{b}{\pi} \left[ \frac{h^2 - (y - y_0) \cdot h}{(x - x_0)^2 + (y - y_0)^2 + h^2} - \arctan\left(\frac{y - y_0}{h}\right) \right]
    \label{y_displ}
\end{equation}

Where $b$ is the Burgers vector, $h$ is the effective depth from the dislocation line, and $x_0, y_0$ are the dislocation positions along x and y, respectively. These equations allow us to compute the displacement field at any arbitrary $(x,y)$ point in the material, considering the influence of dislocations spread across a 2D plane. For the purposes of this model, we assume the dislocations to be randomly placed in the 2D plane as well as non-interacting. An examplary simulated crosshatch pattern is presented in Figure \ref{ml_model}a. We also present the experimental AFM image presented in Fig. \ref{afm}b again, now side-by-side with the 2D version of the simulated crosshatch pattern for comparison. We note good quantitative agreement regarding lambda values, however, the height at the dislocation density presented here is roughly one order of magnitude lower than the experimental data.

\section{Machine Learning Analysis}

When generating the simulated dataset, we simulate the surface 1.5 $\mu$m away from the misfit interface for each dislocation density, consistent with the thickness of the material stack used in our experiments. For simplicity, we assume all dislocations form at the interface with the InP substrate. The simulation was run for 1000 iterations per dislocation density, for 18 different densities ranging from $3 \times 10^{7} /cm^{2}$ to $3 \times 10^{9} /cm^{2}$. With this, we then extract the simulated $\lambda$ and $\epsilon$ values for each simulation. The simulated dataset is plotted in Fig. \ref{ml_model}b,c for $\overline{\lambda}_{x,y}$ and Fig. \ref{ml_model}d,e $\overline{\epsilon}_{x,y}$, respectively. 

\begin{figure}[h!]
    \centering
    \includegraphics[width=\linewidth]{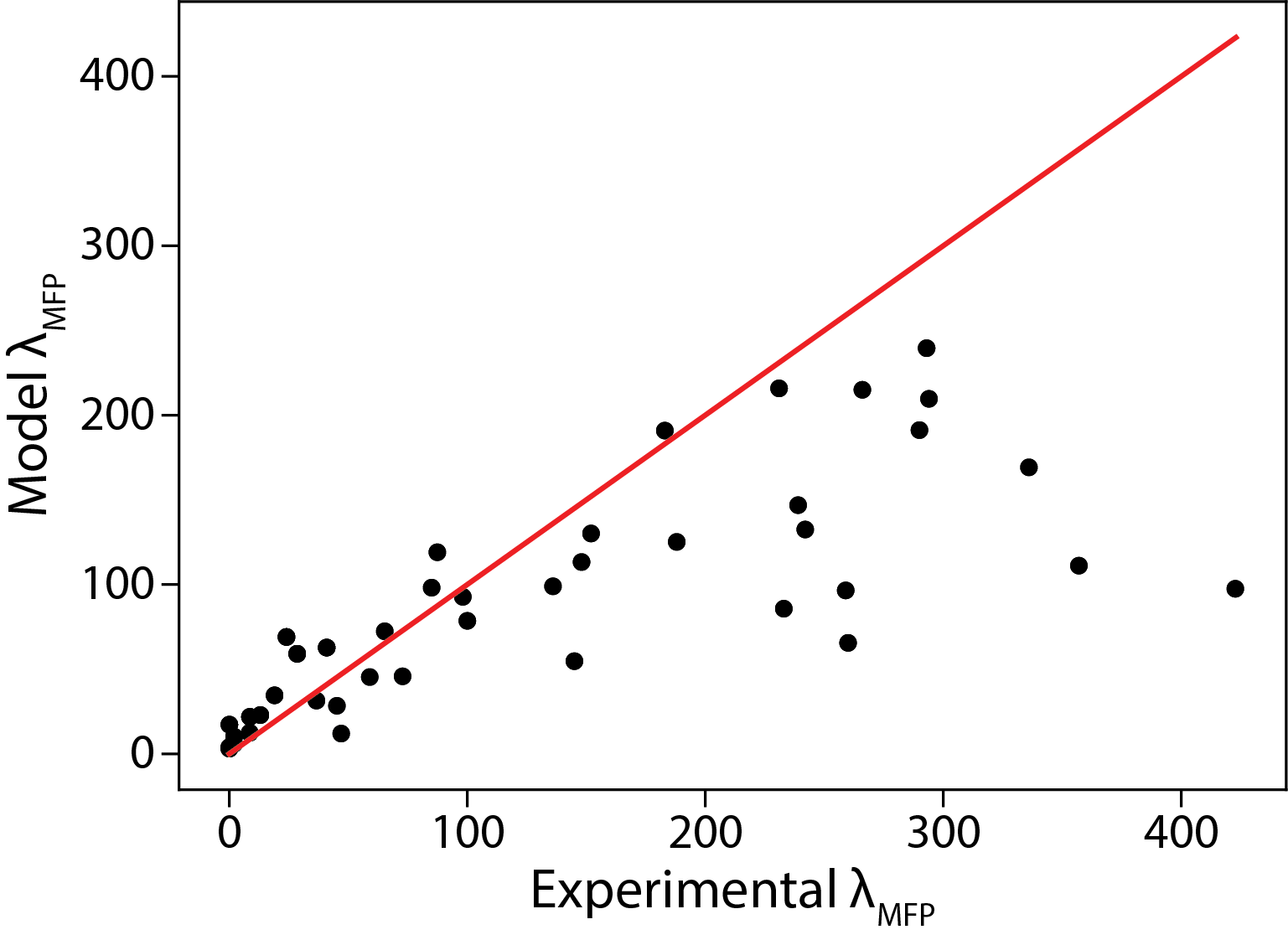}
    \caption{ \textbf{Electron mean free path model.} Machine learning random forest model fitting the experimental electron mean free path to the dislocation model generated in this study. The red line represents perfect agreement between the machine learning model and the experimentally measured quantities.}
    \label{mfp_model}
\end{figure}

The parameters extracted from the AFM simulations -- crosshatch wavelength $\overline{\lambda}$ and amplitude $\overline{\epsilon}$ -- are calculated the same as our previous analysis (schematic shown in Fig. \ref{afm}c). We used these parameters to train a Random Forest machine learning model, which predicted the dislocation density from the AFM-derived $\overline{\lambda}$ and $\overline{\epsilon}$ values. The best hyperparameters for the model were determined through cross-validation and yielded a test $R^2$ value of 0.976, indicating our model captures nearly all of the observed variance in predicting dislocation density. More details about the model can be found in the supplemental materials. 

We then take the model that defines $\rho_{disloc.}(\overline{\lambda},\overline{\epsilon})$ and use it to correlate the experimentally measured values of $\overline{\lambda}$ and $\overline{\epsilon}$ in order to evaluate the dislocation density of our experimentally analyzed AFM surfaces. To evaluate the effectiveness of our model in evaluating the experimental dataset, we predict the dislocation densities for each of the 36 original samples. An example of this comparison is presented in Table \ref{disloc_table}. 

\begin{table}[ht!]
    \centering
    \begin{tabular}{c | c | c | c}
        \textbf{Metric} & \textbf{Simulated} & \textbf{Experimental} & \textbf{Deviation} \\
         \hline
         $R_{a}$ & 0.94 \r{A} & 1.02 \r{A} & 7.8\% \\
        Mean Height & 6.92 \r{A} & 7.10 \r{A} & 2.5\% \\
        $\epsilon$ & 1.21 \r{A} & 1.28 \r{A} & 5.5\%
    \end{tabular}
    \caption{Comparison of simulated crosshatch parameters and experimental crosshatch parameters for a dislocation density of $5.97 \times 10^{8} cm^{-2}$. The experimental dislocation density was determined by the model trained on the simulated dataset.}
    \label{disloc_table}
\end{table}

Where $R_{a}$ is the calculated surface roughness and the mean height is the average valley-peak height for the crosshatch pattern. We note that while our model is rudimentary in the assumption of only one mismatched interface, we find that it suitably captures the behavior of the real system. 

The experimental $\rho_{disloc.}^{exp.}$ values are then correlated with the experimental electron mean free path values, $\overline{\lambda}_{MFP}$, in order to generate a predictive model for $\overline{\lambda}_{MFP}$. The results of this model are presented in Fig. \ref{mfp_model}.

The divergence near 200~nm from the model predictions that indicates the physics observed at larger mean free path values is not accurately captured by our model. Rather, the dependence of $\overline{\lambda}_{MFP}$ on $\overline{\lambda}_{x,y}, \overline{\epsilon}_{x,y}$, and  $\rho_{disloc.}$ is not strongly defined. We emphasize here that the model only accounts for information regarding the structure of the material, and does not consider any form of electronic scattering. This result indicates that this 200~nm $\overline{\lambda}_{MFP}$ is a crossover point where the electron scattering goes from being dominated by dislocation scattering induced by elastic strain and plastic deformation caused by epitaxial growth, to being dominated by ionized impurity and electron-electron scattering. 

\section{Conclusion}

In conclusion, we present here a new method for AFM image analysis using computer image processing to quantitatively parametrize AFM crosshatch patterns and correlate the extracted parametrization with material quality. Using computer vision thresholding, we deconvolute crosshatch images into separate X- and Y- channels and calculate average crosshatch wavelengths, $\overline{\lambda}_{x,y}$, and standard deviations, $\sigma^{\lambda}_{x,y}$, for both channels. AFM height data is used to calculate the average height variation, $\overline{\epsilon}_{x,y}$, and the standard deviations, $\sigma^{\epsilon}_{x,y}$ for both channels. We present newly developed simulation tools to generate AFM crosshatch patterns from a known dislocation density. Using this simulation, we are able to generate large datasets with which to train a machine learning model that relates the crosshatch $\overline{\lambda}$ and $\overline{\epsilon}$ values to an initial dislocation density. From this model we relate the experimental AFM image to a predicted dislocation density in order to correlate electron mean free path, $\overline{\lambda}_{MFP}$, to a dislocation density, $\rho_{disloc}$. We finally present a model based on predicted dislocation density that attempts to predict $\overline{\lambda}_{MFP}$ based solely on strain-induced displacement fields. We observe a divergence between the experimental values and predicted values near $\overline{\lambda}_{MFP}$ = 200 nm that points to this point as a crossover between transport that is dominated by scattering induced by strain-fields and ionized impurity scattering. \\

\section{Experimental Details}

\textit{Material Growth:} InAs near-surface quantum wells are grown in a Varian Gen II molecular beam epitaxy (MBE) system on 50.8 mm InP (001) wafers. The InAs quantum well structures are grown on a strain-relaxed graded buffer layer with top-layer composition of In$_{0.81}$Al$_{0.19}$As. The quantum well is grown to be 4 nm thick and is separated from the wafer surface by a 10 nm thick In$_{0.81}$Ga$_{0.19}$As barrier layer. More details and specifics about this structure growth can be found in reference \cite{strickland2022inasqw}. 

\textit{Atomic Force Microscopy Imaging:} Atomic force microscopy imaging is done with a NanoMagnetics Instruments AFM operating in a dynamical tapping mode. Preliminary AFM image normalization and processing is done using the Gwyddion software package \cite{gwyddion}. 

\textit{Quantum Hall Characterization:} Quantum Hall transport is measured in an Oxford TelsatronPT 1.5 K He4 cryostat. The 2-in wafers are diced into roughly 5$\times$5 mm pieces which then are etched in Transene Type-D (Transene Company) aluminum etchant to remove the top Al layer. Samples are then cleaned in solvents, rinsed in de-ionized water, and then blow-dried with dry nitrogen. Measurements are done in a Van der Pauw wiring configuration.

\section{Supplemental Information}
The supplemental material document contains further details regarding the linear regression model used and the machine learning model data curation, model selection, and training.

\begin{acknowledgments}
We thank Stefano Martiniani for fruitful discussions. The authors would like to acknowledge support for this project from the Office of Naval Research award number N00014-22-1-2764.
\end{acknowledgments}

\section*{Data Availability Statement}

The data that support the findings of this study are available from the corresponding author upon reasonable request. The AFM analysis code is available on github: https://github.com/ShabaniLab.

\bibliography{bibliography}

\end{document}


\preprint{AIP/123-QED}

\title{Supplemental Materials: Machine learning analysis of structural data to predict electronic properties in near-surface InAs quantum wells}
\title{Machine learning analysis of structural data to predict electronic properties in near-surface InAs quantum wells}
\author{Patrick J. Strohbeen}
    \thanks{These authors contributed equally to this work}
\author{Abtin Abbaspour}
    \thanks{These authors contributed equally to this work}
\author{Amara Keita}
    \thanks{These authors contributed equally to this work}
\author{Tarek Nabih}
    \thanks{These authors contributed equally to this work}
\author{Aliona Lejuste}
\author{Andrea Maiani}
    \thanks{Now at NORDITA, Stockholm University}
\author{\\ Alisa Danilenko}
\author{Ido Levy}
\author{Jacob Issokson}
\author{Tyler Cowan}
\author{William M. Strickland}
\author{Mehdi Hatefipour}
\author{Ashley Argueta}
\author{Lukas Baker}
\author{Melissa Mikalsen}
\author{Javad Shabani}
\affiliation{ 
Center for Quantum Information Physics, Department of Physics, New York University, New York, NY 10003 USA
}%

\date{\today}

\maketitle

\section{Linear Regression Model}

\begin{equation}
\begin{split}
    \overline{\lambda}_{MFP} & = 60.59 - 0.01\overline{\lambda}_{x} - 0.05\overline{\lambda}_{y} - 35.98\overline{\epsilon}_{x} - 50.01\overline{\epsilon}_{y} \\
    &+ 0.39\sigma^{\lambda}_{x}- 0.29\sigma^{\lambda}_{y} + 169.72\sigma^{\epsilon}_{x} + 35.59\sigma^{\epsilon}_{y}
    \label{model-1}
\end{split}
\end{equation}

\begin{equation}
\begin{split}
    \overline{\lambda}_{MFP} & = 57.76 - 33.45\overline{\epsilon}_{x} - 49.2\overline{\epsilon}_{y} + 0.4\sigma^{\lambda}_{x} - 0.39\sigma^{\lambda}_{y} \\
    & + 168.02\sigma^{\epsilon}_{x} + 15.59\sigma^{\epsilon}_{y}
    \label{model-2}
\end{split}
\end{equation}

\section{Machine Learning Modeling Process}

The development of our model aimed to predict dislocation density in semiconductor materials based on parameters derived from AFM crosshatch analysis. We utilized a comprehensive dataset containing various features related to the crosshatch pattern, including the mean ($\overline{\lambda}, \overline{\epsilon}$) and standard deviation ($\sigma_{\lambda}, \sigma_{\epsilon}$) of the wavelength ($\lambda$) and amplitude ($\epsilon$) in both the X and Y directions. This section outlines the methodology employed to build, validate, and apply machine learning models for this purpose.

\subsection{Data Preprocessing}

The dataset was first loaded and randomized to ensure unbiased training and testing. Randomization is necessary due to the generation of crosshatch simulation was done in order of increasing dislocation densities. Furthermore, since the features span different scales, it is crucial to standardize them to enhance model performance. We employed the \texttt{StandardScaler} from the Python package \texttt{sklearn} \cite{scikit-learn} to scale all input features. Lastly, the dislocation density values are log-transformed using \texttt{np.log1p}. This process transforms the data into an approximate normal distribution, reducing the variance of the dataset while preserving the magnitude of change across the entries. This ultimately allows for the creation of a more robust model.

\begin{lstlisting}
features = df.columns[:-1]  # Assuming the last column is the target
scaler = StandardScaler()
df[features] = scaler.fit_transform(df[features])
df['log_dislocation_density'] = np.log1p(df['dislocation density'])
\end{lstlisting}

\subsection{Model Selection and Training}

Four machine learning models were considered: Random Forest, Gradient Boosting, Support Vector Machine (SVM), and K-Nearest Neighbors (KNN). Each model was fine-tuned using a grid search approach to identify the optimal hyperparameters that minimize the Mean Squared Error (MSE) while maximizing the $R^2$ score on the validation set.

\begin{lstlisting}
models = {
    "Random Forest": (RandomForestRegressor(), param_grid_rf),
    "Gradient Boosting": (GradientBoostingRegressor(), param_grid_gb),
    "Support Vector Machine": (SVR(), param_grid_svr),
    "K-Nearest Neighbors": (KNeighborsRegressor(), param_grid_knn)
}
\end{lstlisting}

The dataset was split into training and testing sets, with 80\% of the data allocated for training and 20\% for testing. The grid search was conducted using 5-fold cross-validation to ensure that the model's performance was evaluated on multiple subsets of the data.

\begin{lstlisting}
X_train, X_test, y_train, y_test = train_test_split(X, y, test_size=0.2, random_state=42)
\end{lstlisting}

\subsection{Model Evaluation}

The performance of each model was evaluated based on the Mean Squared Error (MSE) and $R^2$ score. The results indicated that the Gradient Boosting model with the parameters \texttt{\{'learning\_rate': 0.1, 'max\_depth': 5, 'n\_estimators': 200\}} achieved the best performance with an MSE of 0.0191 and an $R^2$ score of 0.976, closely followed by the Random Forest and SVM models. The full results of the training are presented in Table \ref{train_table} below.

\begin{table}[h!]
    \centering
    \begin{tabular}{c || c | c }
        Model  & \thead{Mean Squared Error \\ (MSE)} & $R^2$ \\
        \hline
        \hline
        Random Forest  & 0.0197 & 0.975 \\
        Gradient Boosting & 0.0191 & 0.976 \\
        \makecell{Support Vector Machine \\ (SVM)} & 0.0192 & 0.976 \\
        \makecell{K-Nearest Neighbors \\ (KNN)} & 0.0227 & 0.972 
    \end{tabular}
    \caption{Caption}
    \label{train_table}
\end{table}

\begin{lstlisting}
results = {}
for name, (model, param_grid) in models.items():
    grid_search = GridSearchCV(estimator=model, param_grid=param_grid, cv=5, n_jobs=-1, scoring='r2')
    grid_search.fit(X_train, y_train)
    best_model = grid_search.best_estimator_
    
    y_pred = best_model.predict(X_test)
    
    mse = mean_squared_error(y_test, y_pred)
    r2 = r2_score(y_test, y_pred)
    results[name] = {'Best Params': grid_search.best_params_, 'MSE': mse, 'R^2': r2}
\end{lstlisting}

\subsection{Application of the Model}

Since it performed best at fitting the dataset, the Gradient Boosting model was selected as the final model for predicting dislocation density. This model was applied to the experimental dataset containing crosshatch parameters from real analyzed AFM images. The predictions were made by scaling the input features, applying the model to predict the log-transformed dislocation density, and then applying the inverse log transformation to recover dislocation density values corresponding to the experimental dataset.

\begin{lstlisting}
model = joblib.load('gradient_boosting_model.pkl')
scaler = joblib.load('scaler.pkl')

def predict_dislocation_density(lambda_mean_x, lambda_std_x, lambda_mean_y, lambda_std_y,
                                epsilon_x, epsilon_std_x, epsilon_y, epsilon_std_y):
    input_data = pd.DataFrame({
        'lambda_mean_x': [lambda_mean_x],
        'lambda_std_x': [lambda_std_x],
        'lambda_mean_y': [lambda_mean_y],
        'lambda_std_y': [lambda_std_y],
        'epsilon_x': [epsilon_x],
        'epsilon_std_x': [epsilon_std_x],
        'epsilon_y': [epsilon_y],
        'epsilon_std_y': [epsilon_std_y]
    })
    
    scaled_data = scaler.transform(input_data)
    log_dislocation_density_pred = model.predict(scaled_data)
    dislocation_density_pred = np.expm1(log_dislocation_density_pred)
    
    return dislocation_density_pred[0]
\end{lstlisting}

The new dataset, now augmented with the predicted dislocation densities, is saved and utilized to generate the second model for electron mean free path, $\lambda_{MFP}$. The model for $\lambda_{MFP}$ is a Random Forest model, generated using the same procedures.